\documentclass[11pt,showpacs,preprintnumbers,amsmath,amssymb]{revtex4}

\usepackage{graphicx}
\usepackage{dcolumn}
\usepackage{bm}

\begin{document}

\preprint{}

\title{the twist-3 distribution amplitudes in the $B\rightarrow\pi$ transition form factor
\footnote{Supported by National Science Foundation of China.} }
\author{ Ming-Zhen Zhou$^2$}
 \email{zhoumz@mail.ihep.ac.cn}
\author{ Xing-Hua Wu$^2$}
 \email{xhwu@mail.ihep.ac.cn}
\author{Tao Huang$^{1,2}$}
\affiliation{$^1$CCAST(World Laboratory), P.O.Box 8730,
 Beijing 100080, P.R.China}
\affiliation{$^2$Institute of High Energy Physics, P.O.Box 918 ,
Beijing 100039, P.R.China}
\date{\today}

\begin{abstract}
We derive an expression for the $B\rightarrow\pi$ transition form
factor only depending the twist-3 distribution amplitudes by
choosing an adequate chiral current correlator in the light-cone
QCD sum rules. Our result show that the contribution from the
twist-3 distribution amplitudes to the $f_{B\pi}^+(q^2)$ give a
constraint on the twist-3 light-cone distribution amplitude.
\end{abstract}

\maketitle


Heavy-to-light exclusive decays are important for understanding
and testing the standard model, and it is of crucial interest to
make a reliable prediction for these exclusive processes.
Theoretically, the precise calculations of heavy-to-light from
factors are of a great importance. Especially, it will be helpful
for a clear understanding of $B\rightarrow\pi l\bar{v_l}(l=e,\mu)$
which provides us with a good chance to extract the CKM matrix
element $|V_{ub}|$ from the available data. Recent progress on QCD
factorization formula \cite{Beneke}, which was proposed for
$B\rightarrow\pi\pi$, $\pi$K and $\pi D$, show that the amplitudes
for these nonleptonic decays can be expressed in terms of the
semileptonic form factors, hadronic light-cone distribution
amplitudes and hard-scattering functions that are calculated in
perturbative QCD(pQCD). For the semileptonic form factors, one can
take them as inputs from experimental data directly.

Many papers have tried to confront calculations of the
semileptonic form factors. For example, these form factor can be
calculated by pQCD \cite{Li} and by applying the light-cone QCD
sum rules \cite{Belyaev,huang}. In fact, a considerable
long-distance contribution may dominate the heavy-light factors.
The pQCD approach adapts the modified hard-scattering amplitude to
them by a resummation of Sudakov logarithms, which can suppress
the soft contribution beyond naive power counting. In light-cone
QCD sum rules, the contribution of nonperturbative dynamics are
attributed to the distribution amplitudes which are classified by
their twists.

The $B\rightarrow\pi$ transition form factor were calculated in
the light-cone QCD sum rules \cite{Belyaev}. Remarkably, the main
uncertainties in these calculations arise from light-cone
distribution amplitudes which include not only the twist-2
distribution amplitude but the twist-3 and the twist-4
distribution amplitudes. The later two distribution amplitudes are
understood poorly. It was shown that the contribution of the
twist-3 distribution amplitudes is about 30-50$\%$ and the
contribution of the twist-4 distribution amplitudes is about 5$\%$
to $B\rightarrow\pi$ transition form factor. Thus the great
uncertainty, if possible, would be due to the uncertainty in the
twist-3 distribution amplitudes in the framework of the light-cone
QCD sum rules. In order to reduce the uncertainty Ref.
\cite{huang} takes an adequate chiral current correlator to make
the contribution of the twist-3 distribution amplitudes vanish in
the $B\rightarrow\pi$ transition form factor. Consequently, the
possible pollution by them can be avoided in the $B\rightarrow\pi$
transition form factors.

It is very interesting to ask a similar question if one can derive
an expression for the $B\rightarrow\pi$ transition form factor
only depending the twist-3 distribution amplitudes by choosing the
chiral current correlator. The answer is positive. We will discuss
the question in this paper.

Let us start with the following definition of $B \rightarrow \pi$
transition form factor $f_{B \pi}^{+}(q^2)$,
\begin{equation}
\langle \pi(p) | \bar{u} \gamma_{\mu} b | B(p+q) \rangle = 2 f_{B
\pi}^+(q^2)p_{\mu}+f_{B \pi}^-(q^2)q_{\mu}
\end{equation}
with $q$ being the momentum transfer. In order to calculate the
form factor we need to construct a correlator. The different
correlator gives the different expression (see Ref. \cite{Belyaev}
and Ref. \cite{huang}). For example, Ref.\cite{huang} proposed an
improved approach by choosing the chiral current and they got the
transition form factor,
\begin{eqnarray}
f_{B \pi}^+(q^2) & & = \frac{m_0^2 f_{\pi}}{m_B^2 f_B}
e^{m_B^2/M^2}
      \left\{ \int_{\Delta}^1 \frac{du}{u} exp \left[-\frac{m_b^2-q^2(1-u)}{u M^2} \right]
      \cdot \left[\phi_{\pi}(u)-\frac{4m_b^2}{u^2 M^4}g_1(u)
      \right. \right. \nonumber \\
& &   \left. \left. +\frac{2}{u M^2}\int_0^u dv g_2(v)
      \left( 1 +\frac{m_b^2+q^2}{u M^2} \right) \right]
      + \int_0^1 dv \int D \alpha_i\frac{\theta(\alpha_1 + v \alpha_3
      -\Delta)} {(\alpha_1 + v \alpha_3)^2 M^2} \right. \nonumber \\
& &   \left. exp\left[- \frac{m_b^2-(1-\alpha_1 - v
      \alpha_3)q^2}{M^2
      (\alpha_1 + v \alpha_3)} \right]
      \left[ 2 \phi_{\bot}(\alpha_i)+2 \widetilde{\phi}_{\bot}(\alpha_i)
      - \phi_{\|}(\alpha_i) - \widetilde{\phi}_{\|}(\alpha_i)\right]
      \right. \nonumber \\
& &   \left. -4 m_b^2 e^{-s_0/M^2} \left[ \frac{1}{(m_b^2-q^2)^2}
      \left( 1 + \frac{s_0-q^2}{M^2} \right)
      g_1(\Delta) -\frac{1}{(s_0-q^2)(m_b^2-q^2)\frac{d g_1(\Delta)}{d u}}  \right]
      \right. \nonumber \\
& &   \left. -2
      e^{-s_0/M^2}\left[\frac{m_b^2+q^2}{(s_0-q^2)(m_b^2-q^2)}g_2(\Delta)-
      \frac{1}{m_b^2-q^2} \left( 1+\frac{m_b^2+q^2}{m_b^2-q^2} \right)
      \left( 1 + \frac{s_0-q^2}{M^2}\right) \right.\right.
      \nonumber\\
& &   \left.\left. \int_0^{\Delta} dv g_2(v)\right] \right\}
\end{eqnarray}
with $ D \alpha_{i}=d \alpha_1 d \alpha_2 d \alpha_3
\delta(1-\alpha_1-\alpha_2-\alpha_3) $,
$m_0=\frac{m_\pi^2}{m_u+m_d}$ and $ \Delta=(m_{b}^2-q^2)/(s_0-q^2)
$. Here $ \phi_\pi $ is $ \pi $ meson twist-2 distribution
amplitude, $
g_1(u),g_2(u),\phi_{\bot}(\alpha_i),\widetilde{\phi}_{\bot}(\alpha_i),
\phi_{\|}(\alpha_i),\widetilde{\phi}_{\|}(\alpha_i)$ are $ \pi $
meson twist-4 distribution amplitudes$, s_0$ is the threshold
parameter which should be set to the value near the squared mass
of the lowest scalar $B^*$ meson, and $M$ is the Borel parameter.

Now we propose to chose another chiral current to construct a
correlator,
\begin{eqnarray}
\Pi_{\mu}(p,q) & & = i \int d^4 x e^{i q \cdot x} \langle \pi(p) |
T \left\{ \bar{u}(x)\gamma_{\mu} (1+\gamma_5) b(x) ; \bar{b}(0)i m_b (1-\gamma_5) d(0) \right\} | 0 \rangle \nonumber \\
& & = \Pi(q^2,(p+q)^2)p_{\mu} +
\widetilde{\Pi}(q^2,(p+q)^2)q_{\mu},
\end{eqnarray}
which is different from that in Ref.\cite{huang}. Here the chiral
limit $ p^2=m_{\pi}^2=0 $ is made.

This correlator can be calculated in two ways. First, we discuss
the hadronic representation for the correlator by inserting a
complete series of intermediate state with the same quantum number
as the current operator $ \bar{b} ~ i(1-\gamma_5) d $ in it. Then
isolating the pole term of lowest pseudoscalar $ B $ meson, we get
the result,
\begin{eqnarray}
\Pi_{\mu}^H(p,q) & & =\Pi^H(q^2,(p+q)^2)p^{\mu}+\widetilde{\Pi}^H(q^2,(p+q)^2)q_{\mu} \nonumber \\
& & = - \frac{ m_b \langle \pi | \bar{u}\gamma_{\mu}b| B \rangle
\langle B | \bar{b}\gamma_5 d | 0 \rangle}{m_B^2-(p+q)^2} +
\Sigma_{H} \frac{ m_b \langle \pi |
\bar{u}\gamma_{\mu(1+\gamma_5)}b| B_H \rangle \langle B_H |
\bar{b}(1-\gamma_5) d | 0 \rangle}{m_{B_H}^2-(p+q)^2}.
\end{eqnarray}
Here the intermediate states $ B_{H} $ contain not only
pseudoscalar resonances of the masses greater than $ m_B $, but
also scalar resonance with $ J^P=0^+ $, corresponding to the
operator $ \bar{b}d $. Substituting Eq.(1) and the definition $
m_b \langle B| \bar{b}i\gamma_5 d| 0 \rangle =m_B^2 f_B $ into
Eq.(4), the invariant amplitudes $ \Pi^H $ and $ \widetilde{\Pi}^H
$ become
\begin{equation}
\Pi^H[q^2,(q+p)^2]=\frac{-2 f_{B\pi}^+ m_B^2 f_B}
{m_B^2-(p+q)^2}+\int_{s_0}^{\infty}\frac{\rho^H(s)} {s-(p+q)^2}
ds+subtraction
\end{equation}
and
\begin{equation}
\tilde{\Pi}^H[q^2,(q+p)^2]=\frac{-f_{B\pi}^- m_B^2 f_B}
{m_B^2-(p+q)^2}+\int_{s_0}^{\infty}\frac{\tilde{\rho}^H(s)}
{s-(p+q)^2} ds+subtraction.
\end{equation}
The terms in the integration are the contribution from higher
resonances and continuum states above threshold $ s_0 $. Due to
the quark-hadron duality ansatz, the spectral densities $
\rho^H(s) $ and $ \widetilde{\rho}^H(s) $ can be approximated by
the following expression,
\begin{equation}
\rho^H(s)=\rho^{QCD}(s)\theta(s-s_0)\quad and \quad
\tilde{\rho}^H(s)=\tilde{\rho}^{QCD}(s)\theta(s-s_0).
\end{equation}
On the other hand, the correlator can be calculated in QCD theory,
to obtain the desired sum rule for $ f_{B \pi}^+(q^2) $, we work
in the large space-like momentum regions $ (p+q)^2-m^2_b \ll 0  $
for the $ b \bar{d} $ channel, and $ q^2 \ll m_b^2-0(1GeV^2) $ for
the momentum transfer, which correspond to the light cone region $
x^2 \simeq 0 $ and are required by the validity of the operator
product expansion(OPE). First we can write down a full $b$-quark
propagator,
\begin{eqnarray}
\langle 0\mid T{b(x)\bar{b}(0)}\mid0 \rangle & = & i \int
\frac{d^4k} {2 \pi}^4
e^{ik\cdot x } \frac{\hat{k}+m_b} {k^2-m_b^2}-ig_s \int \frac{d^4k} {2 \pi}^4 e^{i k\cdot x}  \nonumber \\
& & \int_{0}^{1}dv \left[ \frac{1} {2} \frac{\hat{k}+m_b}
{(m_b^2-k^2)^2} G^{\mu\nu}(vx)\sigma_{\mu\nu}+\frac{1}
{m_b^2-k^2}vx_\mu G^{\mu\nu}(vx)\gamma_\nu \right],
\end{eqnarray}
where $G_{\mu \nu}$is the gluonic field strength and $ g_s $
denote the strong-coupling constant. Carrying out the OPE for the
correction and making use of Eq.(8), we require several formulas
in Ref.\cite{Braun}
\begin{eqnarray}
& & \langle \pi(q)\mid \bar{u}(x)i\gamma_5d(0)\mid0 \rangle
=\frac{f_\pi
m_\pi^2}{m_u+m_d} \int_{0}^{1}du e^{iu q\cdot x} \phi_p(u) \nonumber \\
& & \langle\pi(q)\mid \bar{u}(x)\sigma_{\mu\nu}\gamma_5d(0) \mid 0
\rangle=i (q_\mu x_\nu -q_\nu x_\mu)\frac{f_\pi m_\pi^2}
{6(m_u+m_d)}\int_{0}^{1}du e^{i uq\cdot x} \phi_\sigma(u) \nonumber \\
& & \langle \pi(q)\mid \bar{u}(x)g_s
G_{\mu\nu}(vx)\sigma_{\alpha\beta}\gamma_5d(0) \mid 0 \rangle
\nonumber \\
& & \quad\quad\quad\quad\quad=i f_{3 \pi}[(q_\mu q_\alpha
g_{\nu\beta}-q_\nu q_\alpha
g_{\mu\beta})-(\alpha\leftrightarrow\beta)]\int D\alpha_i \phi_{3
\pi}(\alpha_i) e^{iq\cdot x(\alpha_1+v \alpha_3)}.
\end{eqnarray}
Here $ \phi_p(u) $ and $ \phi_\sigma(u) $ are twist-3 distribution
amplitudes of $ \pi  $ meson, $ \phi_{3 \pi}(\alpha_i) $ is
twist-3 three-particle distribution amplitude of $ \pi  $ meson.
Substituting the above $b$-quark full propagator and the
corresponding $ \pi $ meson distribution amplitudes into the
correlator and completing the integrations over $ x $ and $ k $
variables, finally we obtain an expression,
\begin{eqnarray}
\Pi^{QCD}[q^2,(p+q)^2] & = & \frac{-2 f_\pi m_\pi^2} {m_u+m_d}
\int_{0}^{1}du \frac{1} {m_b^2-(p+uq)^2} \left[
u\phi_p(u)+\frac{1} {6} (2+\frac{p^2+m_b^2}
{m_b^2-(p+uq)^2})\phi_\sigma(u) \right]
\nonumber \\
& &  +\int_{0}^{1}dv\int D\alpha_i
\frac{-8f_{3\pi}\phi_{3\pi}(\alpha_i)vq\cdot p}
{(m_b^2-(p+(\alpha_1+v\alpha_3)q)^2)^2}
\end{eqnarray}
After substituting Eq.(10) into Eq.(5) and performing the Borel
transformation with respect to $ (p+q)^2 $, a sum rule for the
$B\rightarrow\pi$ transition form factor can be obtained
\begin{eqnarray}
f_{B\pi}^+(q^2) & = & \frac{m_b f_\pi} {m_B^2 f_B} \frac{m_\pi^2}
{m_u+m_d}exp[\frac{m_B^2} {M^2}] \left \{ \int_{\Delta}^{1}du
exp[-\frac{m_b^2-q^2(1-u)} {u M^2}] [u \phi_p(u)+\frac{1}
{6}(2+\frac{m_b^2+q^2} {u
M^2})\phi_\sigma(u)] \right. \nonumber \\
& & \left. -\frac{2f_{3\pi}} {f_\pi} \frac{m_u+m_d}
{m_\pi^2}\int_{0}^{1}dv \int D\alpha_i \frac{\theta(\alpha_1+v
\alpha_3-\Delta)} {(\alpha_1+v
\alpha_3)^2}exp[-\frac{m_b^2-q^2(1-\alpha_1-v \alpha_3)}
{(\alpha_1+v \alpha_3)M^2}]\right. \nonumber \\
& & \left. \cdot[1-\frac{m_b^2-q^2} {(\alpha_1+v
\alpha_3)M^2}]\phi_{3\pi}(\alpha_i) \right \},
\end{eqnarray}
where M is the Borel parameter. Eq.(11) shows that
$f_{B\pi}^+(q^2)$ only depends on the twist-3 distribution
amplitudes. It means that the contribution from the twist-3
distribution amplitudes to the $f_{B\pi}^+(q^2)$ has the same
order of magnitude as that from the leading twist distribution
amplitude.

Now we need to make a choice of input parameters entering the sum
rule Eq.(11) for $f_{B\pi}^+(q^2)$. Let us specify the twist-3
model of the pion distribution amplitudes, $\phi_p(u)$
,$\phi_\sigma(u)$ and
$\phi_{3\pi}(\alpha_i)$(Ref.\cite{Belyaev,Braun}),
\begin{equation*}
\phi_p(u)=1+B_2\frac{1}{2}(3(2u-1)^2-1)+B_4\frac{1}{8}(35(2u-1)^4-30(2u-1)^2+3),
\end{equation*}
\begin{equation*}
\phi_\sigma(u)=6u(1-u) \left [
1+C_2\frac{3}{2}(5(2u-1)^2-1)+C_4\frac{15}{8}(21(2u-1)^4-14(2u-1)^2+1)
\right ]
\end{equation*}
and
\begin{eqnarray*}
& & \phi_{3\pi}(\alpha_i)=360\alpha_1\alpha_2\alpha_3^2 \nonumber \\
& &  \quad\quad\quad\quad
               \left[1+\varpi_{1,0}\frac{1}{2}(7\alpha_3-3)
               +\varpi_{2,0}(2-4\alpha_1\alpha_2-8\alpha_3+8\alpha_3^2)
               +\varpi_{1,1}(3\alpha_1\alpha_2-2\alpha_3+3\alpha_3^2)
               \right ],
\end{eqnarray*}
where $B_2=30R$,
$B_4=\frac{3}{2}(4\varpi_{2,0}-\varpi_{1,1}-2\varpi_{1,0})$,
$C_2=R(5-\frac{1}{2}\varpi_{1,0})$,
$C_4=\frac{1}{10}R(4\varpi_{2,0}-\varpi_{1,1})$ with
$R=\frac{f_{3\pi}}{m_0f_\pi}$, $f_\pi=133MeV$,
$f_{3\pi}=0.0026GeV^2$ and $\varpi_{1,0}=-2.18$,
$\varpi_{2,0}=8.12$, $\varpi_{1,1}=-2.59$. Other input parameters
are taken in the following: $s_0=33GeV^2$, $M^2=16GeV^2$,
$m_b=4.7GeV$, $m_B=5.28GeV$ and $f_B=165MeV$. With these inputs,
we ca carry out the numerical analysis. The form factor Eq.(11) in
this paper is depicted by the solid curve in Fig.1. The dashed and
dotted curves in Fig.1 are taken from Ref.\cite{Belyaev} and
Ref.\cite{huang} respectively. It shows that three curves are
consistent in the region $q^2<16GeV^2$. In fact, the applicability
of the light cone QCD sum rules is questionable as
$q^2\geqslant18GeV^2$ \cite{huang}, and a comparison between the
different approaches in the regions is meaningless. Also one can
see from Fig.1 that the form factor go up very quickly beyond the
region $q^2=15GeV^2$ as long as the twist-3 distribution
amplitudes make contribution to sum rules.

In summary, we show that the different expressions for the
$B\rightarrow\pi$ transition form factor by choosing the different
adequate current correlator in the light cone QCD sum rules.
Especially, we derive the expression for $f_{B\pi}^+(q^2)$ only
depending the twist-3 distribution amplitudes. It is consistent
with other expressions by employing the present model for the pion
distribution amplitudes. Conversely, they provide constraints on
the pion distribution amplitudes.

\newpage
\begin{figure}[tbp]
\begin{center}
\includegraphics*[scale=1.3,angle=0.]{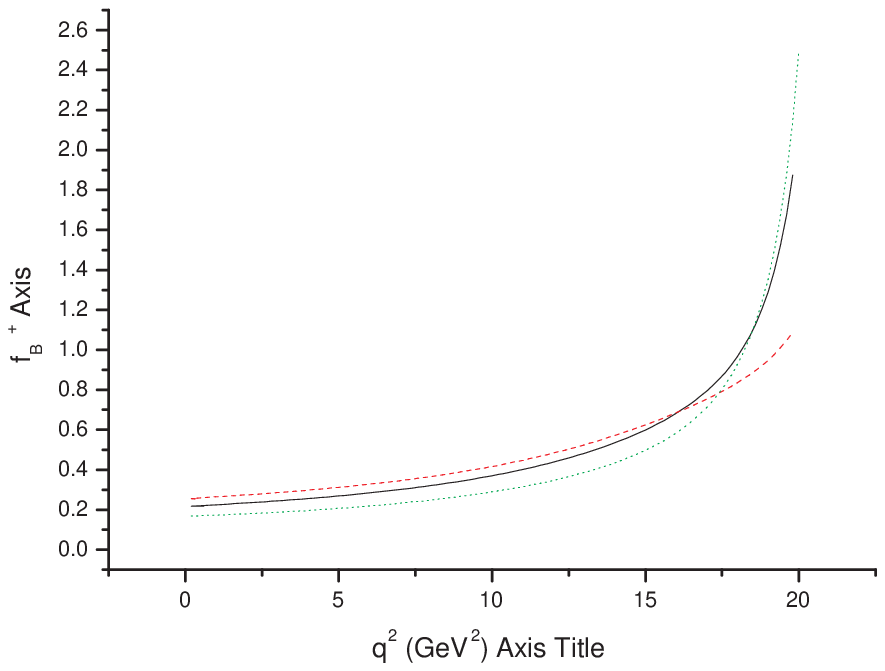}
\end{center}
\caption{The transition form factor $f_{B\pi}^+(q^2)$ in the light
cone QCD sum rules at $M^2=16GeV^2$ with $s_0=33GeV^2$,
$m_b=4.7GeV$, $m_B=5.28GeV$, $f_B=165MeV$, $f_\pi=132MeV$.}
\end{figure}

\end{document}